\documentclass[twocolumn,showpacs,preprintnumbers,amsmath,amssymb,prl]{revtex4}
\usepackage{graphicx}
\usepackage{color}
\usepackage{dcolumn}
\usepackage{bm}

\def \be {\begin{equation}}
\def \ee {\end{equation}}
\def \ben {\begin{eqnarray}}
\def \een {\end{eqnarray}}
\begin{document}
\bibliographystyle{../prsty}

\title{Generalized master equation for modular exciton density transfer}
     
\author{Seogjoo Jang,$^{1}$\footnote{Corresponding Author} Stephan Hoyer,$^2$  Graham Fleming,$^{3,4}$ and K. Birgitta Whaley$^3$}
\affiliation{$^1$Department of Chemistry and Biochemistry, Queens College and the Graduate Center, City University of New York, 65-30 Kissena Blvd., Flushing, NY 11367\\ $^2$Department of Physics, University of California, Berkeley, CA 94720 \\
$^3$Department of Chemistry, University of California, Berkeley, CA 94720 \\
$^4$Physical Biosciences Division, Lawrence Berkeley National Laboratory, Berkeley CA 94720} 
\date{\today}

\begin{abstract}
A generalized master equation (GME) governing quantum evolution of modular exciton density (MED) is derived for large scale light harvesting systems composed of weakly interacting modules of multiple chromophores.  The GME-MED offers a practical framework to incorporate real time coherent quantum dynamics calculations at small length scales into dynamics over large length scales, without assumptions of time scale separation or specific forms of intra-module quantum dynamics.   A test of the GME-MED for four sites of the Fenna-Matthews-Olson complex demonstrates how 
coherent dynamics of excitonic populations over many coupled chromophores can be accurately described
by transitions between subgroups (modules) of delocalized excitons.
\end{abstract}

\maketitle
Many photosynthetic units of bacteria and higher plants have modular structures where the entire systems are composed of smaller subunits, or 
``modules" of protein-chromophore complexes\cite{hu-qrb35,blankenship-ps}.   While the nature of interactions and
quantum dynamics within each module varies, the inter-module interactions are 
generally weak.  A striking characteristic in natural photosynthetic systems is that  excitons can migrate through those weak links and find their destinations with near unit efficiency within picoseconds.
How can this be accomplished  despite significant disorder and fluctuations?  What are the general conditions 
ensuring such high efficiency of natural systems?  Recent theoretical studies provide some clues\cite{jang-jpcb111,rebentrost-njp11,caruso-pra81,wu-njp12}, but the answers for the above fundamental questions are far from being settled.  To this end, quantitative  elucidation of the dynamics 
over larger length and long time  scales is 
needed.  However, accurate quantum dynamical calculations are typically limited to small ($\sim 7$ chromophores) \cite{ishizaki-pnas106,huo-jpcl2}  or medium range systems 
having up to $\sim 30$ chromophores \cite{strumpfer-jcp137,hein-njp14,kreisbeck-jpcl3} 
with the latter already requiring massive computational resources.
Simulation of larger scale complexes with hundreds of chromophores (e.g., photosystem II) using such accurate techniques is impractical.  To date, such simulations 
for larger systems have therefore relied instead on Pauli master equations\cite{ritz-jpcb105,sener-jcp120,yang-bj85,novoderezhkin-pccp13,renger-pr102,renger-jpp168,bennett-jacs}  
but without clear microscopic derivation of the equation or theoretical justification for adopting  particular forms of rate kernels.
This makes it difficult to assess the reliability or to make further improvement of such approaches.   In this work, we derive a generalized master equation (GME) for a coarse-grained exciton density over chromophore subunits or modules. 
This serves as a practical approach to bridge the gap between complex sizes for which accurate quantum dynamics calculations are possible and the demand for simulation of energy transfer over larger length scales in photosynthetic and related complex systems.

The GME for modular exciton density transfer (GME-MED) derived in this work provides a rigorous formulation of a recent work\cite{hoyer-pre86}, where a stochastic description of conditional inter-module transport was proposed with rates accounting for the intra-module quantum coherence.  We show the GME-MED  reproduces  known equations in appropriate limits,  clarifies assumptions underlying the use of multichromophoric F\"{o}rster resonance energy transfer (MC-FRET) rate\cite{scholes-jpcb105,jang-prl92,jang-jpcb111} in a Pauli master equation, and provides a practical means to incorporate high level intra-module quantum calculations into energy transfer simulation over significantly longer length scales.

\begin{figure}
\includegraphics[width=2.8 in]{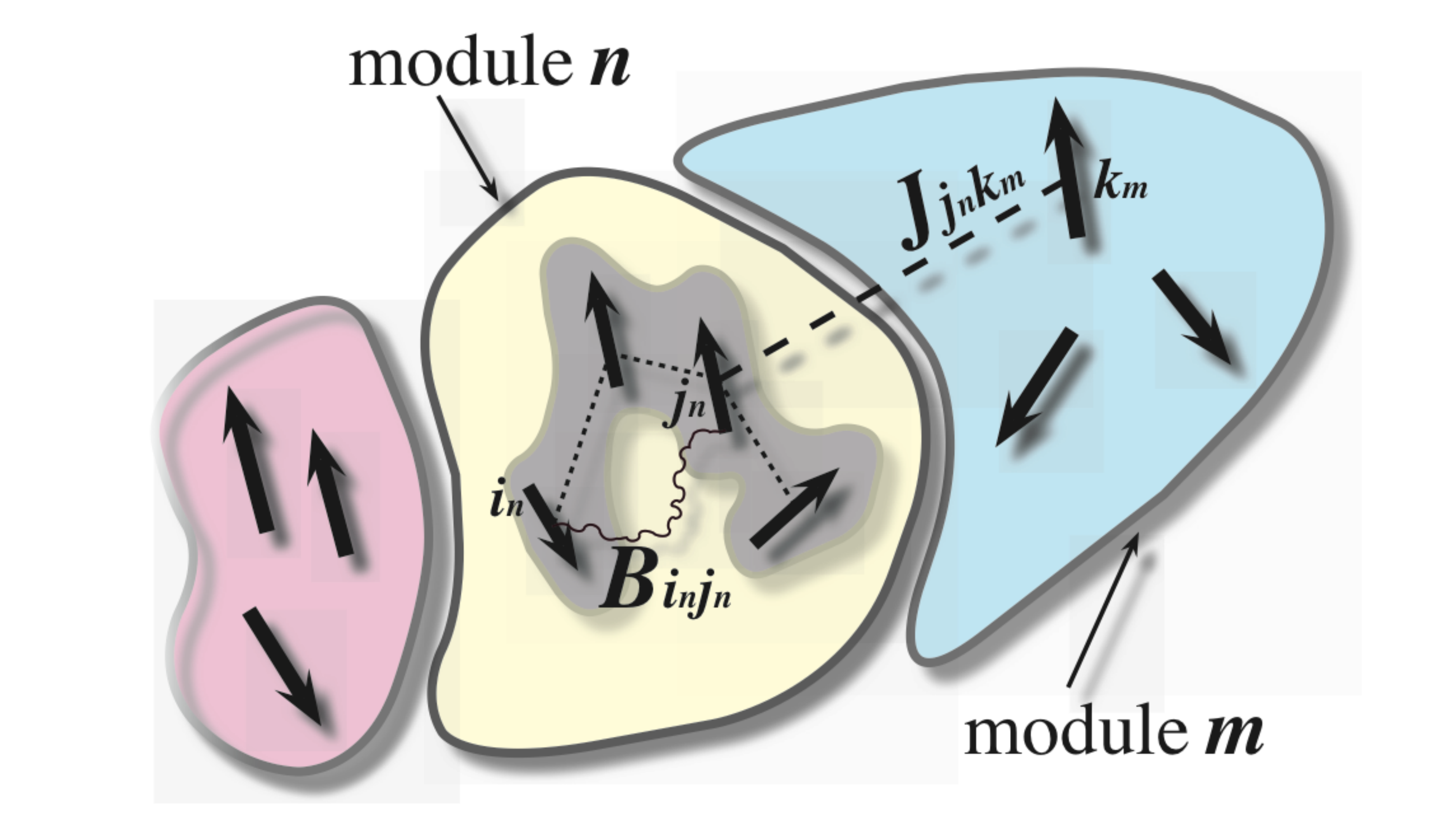} 
\caption{Schematic of a modular system. Arrows represent transition dipoles of each chromophore, dotted lines the electronic coupling, and wavy lines the system-bath coupling.  The grey region represents the modular density of a delocalized exciton. }
\end{figure}

Consider a total Hamiltonian given by $H=H_0+H_c$, where  $H_0$ represents noninteracting modules of excitons together with their environmental degrees of freedom and $H_c$ the couplings between different modules.  Each module is denoted as $n$ or $m$, and a chromophore in the $n$th module is denoted as $j_n$, $k_n$, etc.  Thus, 
\be
H_0=\sum_n H_n =\sum_n 
\{ H_n^e+\sum_{i_n, j_n} B_{i_nj_n} |i_n\rangle\langle j_n|+H_n^g \}\ ,
\ee
with $H_{n}^e$ the single exciton Hamiltonian of the $n$th module, $|i_n\rangle$ 
the site excitation of the $i_n$th chromophore in the $n$th module, $B_{i_nj_n}$ 
the bath operator coupled to the excitonic coupling term $|i_n\rangle\langle j_n|$, and $H_n^g$ 
the bath Hamiltonian (the Hamiltonian in the ground electronic state) of the $n$th module.   
The inter-module coupling Hamiltonian has the form:   
\be
H_c=\sum_{n,m}\sum_{j_n, k_m}J_{j_nk_m} |j_n\rangle\langle k_m| \ , \label{eq:hc_exp}
\ee
where $J_{j_nk_m}$ is assumed to be real and symmetric, and $J_{j_nk_m}=0$ for $n=m$.  For generality, it is assumed that $H_n^e$, $B_{i_nj_n}$,  and $H_c$ can be time dependent although we do not show this explicitly, whereas $H_n^g$ remains time independent.   Figure 1 illustrates an example of a modular structure. 

We denote the time evolution operator for the interaction free Hamiltonian $H_0$ as  $U_0(t,t')=\exp_{(+)}\{-i\int_{t'}^t d\tau H_0(\tau)/\hbar\}=\prod_n U_n(t,t')$, where $U_n(t,t')=\exp_{(+)}\{-i\int_{t'}^t d\tau H_n(\tau)/\hbar\}$ 

with the subscript (+) implying chronological time ordering.  Assuming the exciton is created at time $t=0$, we shall abbreviate
$U_0(t,0)$ and $U_n(t,0)$ as $U_0(t)$ and $U_n(t)$.
The total density operator is denoted as $\rho(t)$.  In the interaction picture with respect to $H_0$, $\rho_I(t)=U_0^\dagger (t) \rho (t) U_0(t)$ evolves according to 
\ben
\frac{\partial \rho_I(t)}{\partial t}=-\frac{i}{\hbar} [H_{c,I}(t),\rho_I(t)]=-i{\mathcal L}_{c,I}(t)\rho_I (t)\ , \label{eq:liouville_int}
\een
where  $H_{c,I}(t)=U_0^\dagger(t) H_c U_0(t)$.  The second equality in Eq. (\ref{eq:liouville_int}) serves as the definition of ${\mathcal L}_{c,I}(t)$.  

The ground state time evolution operator of the $n$th module is denoted as $U_n^g(t)=\exp\{-i t H_n^g/\hbar\}$.  Note that $|j_n\rangle$ represents the state where only the $j_n$th chromophore in the $n$th module is excited while all other modules are in the ground electronic state.  Thus,  $\langle j_n|U_0(t)=\left (\prod_{m\neq n} U_m^g(t) \right ) \langle j_n| U_n(t)$, and
\be
H_{c,I}(t)=\sum_{n,m}\sum_{j_n,k_m} J_{j_nk_m} {\mathcal T}_{j_nk_m}(t)=\sum_{n,m} {\mathcal F}_{nm} (t) \ ,  \label{eq:hc_it}
\ee
where  ${\mathcal T}_{j_nk_m}(t)=U_0^\dagger (t)|j_n\rangle\langle k_m|U_0(t)=U_n^\dagger (t) U_{n}^g(t)|j_n\rangle\langle k_m|U_{m}^{g\dagger }(t)U_m(t)$ and ${\mathcal F}_{nm}(t)=\sum_{j_n,k_m} J_{j_nk_m} {\mathcal T}_{j_nk_m}(t)$.   By definition,  ${\mathcal F}_{nm}(t)$ vanishes for $n=m$.  The identity operator in the single exciton space of each module is defined as $1_n=\sum_{j_n}|j_n\rangle\langle j_n|$, and that in the total
single exciton space is defined as $1=\sum_n 1_n$.   The equilibrium bath canonical density operator of the $n$th module in the ground electronic state is denoted as $\rho_{bn}=e^{-\beta H_n^g}/Tr_b \{ e^{-\beta H_n^g}\}$.  

The key idea in deriving the GME-MED is to introduce the following {\it modular projection} super-operator ${\mathcal P}$:
\be
{\mathcal P} (\cdot) =\sum_n \rho_{bC_n}Tr_{bC_n}\{1_n(\cdot)1_n \} \ , \label{eq:proj-def}
\ee
where $(\cdot)$ represents an arbitrary operator, $\rho_{bC_n}=\prod _{m\neq n}\rho_{bn}$, and $Tr_{bC_n}$ represents the trace over all 
baths except for those associated with the $n$th module.  Physically, ${\mathcal P}$ projects the total density operator into an independent sum of blocks, each representing a module.  This satisfies the required condition of ${\mathcal P}^2={\mathcal P}$.  We assume an initial condition at time $t=0$ with no intermodule quantum coherence,
resembling the conditions created by an incoherent light source.   This implies that $(1-{\mathcal P})\rho_I(0)=0$.  
One can also verify that ${\mathcal P} {\mathcal L}_{c,I}(t) {\mathcal P} =0$.  Then, application of ${\mathcal P}$ to Eq. (\ref{eq:liouville_int}) with standard projection operator techniques\cite{vankampen-jsp87,jang-jcp116} leads to
\ben
&&\frac{\partial}{\partial t}  {\mathcal P}\rho_I(t)=-\int_0^t d\tau {\mathcal P} {\mathcal L}_{c,I}(t)\nonumber \\
&&\hspace{.5in} \times e_{(+)}^{-i\int_\tau^t d\tau' (1-{\mathcal P}) {\mathcal L}_{c,I}(\tau') }{\mathcal L}_{c,I}(\tau) {\mathcal P}\rho_I(\tau) \ . \label{eq:prhoi_dt_ex}
\een
The total $n$th module density operator in the interaction picture is given by $\rho_{n,I}(t)=Tr_{bC_n} \left\{1_n \rho_{I}(t)1_n\right\}=U_n^\dagger(t)1_n Tr_{bC_n}\{\rho(t)\}1_n U_n(t)$.  Thus, application of $Tr_{bC_n}\left\{1_n ( \cdot  ) 1_n\right\}=1_nTr_{bC_n}\{(\cdot)\}1_n$ to Eq. (\ref{eq:prhoi_dt_ex}) results in 
\ben
&&\frac{\partial}{\partial t} \rho_{n,I}(t)=-\sum_m \int_0^t d\tau\  1_n Tr_{bC_n}\bigg \{{\mathcal L}_{c,I}(t) \nonumber \\ 
&& \times e_{(+)}^{-i\int_\tau^t d\tau' (1-{\mathcal P}) {\mathcal L}_{c,I}(\tau') } {\mathcal L}_{c,I}(\tau) \rho_{bC_m}\rho_{m,I}(\tau)  \bigg \}1_n\ ,\label{eq:rho_id-dt_ex}
\een
which is still exact.  Under the assumption that the inter-module coupling $H_c$ is small compared to $H_0$, an approximation of $e_{(+)}^{-i\int_\tau^t d\tau' (1-{\mathcal P}) {\mathcal L}_{c,I}(\tau') }\approx 1$ in Eq. (\ref{eq:rho_id-dt_ex}) leads to the following 2nd order approximation in the coupling $H_c$: 
\ben
&&\frac{\partial}{\partial t} \rho_{n,I}(t)= -\sum_m \int_0^t d\tau  \nonumber \\
&&\hspace{.1in}1_n Tr_{bC_n}\left\{{\mathcal L}_{c,I}(t){\mathcal L}_{c,I}(\tau)  \rho_{bC_m}\rho_{m,I}(\tau) \right\}1_n\ .\label{eq:rho_id-dt1}
\een
Note we have made no assumption of weak chromophore-environment coupling (recall that $H_0$ is the sum of the intra-module Hamiltonians together with their environmental couplings).  Eq.~(\ref{eq:rho_id-dt1}) is thus distinct from previous well-known second order expressions for excitonic energy transfer in light harvesting systems \cite{renger-pr343}. 
It provides a complete prescription to incorporate full quantum dynamics calculations for each module (made using, e.g., the methods of\cite{miller-jpca105,makri-arpc50}) into a consistent description of the dynamics across all coupled modules.  The only assumption invoked here is the smallness of  $H_c$ compared to $H_0$: even if there is no natural division into modules, this condition can always be satisfied by choosing a large enough module size. 

When the main focus is on the exciton states, the equation for the reduced system density operator, $\sigma_{n,I}(t)=Tr_{bn}\{\rho_{n,I}(t)\}$, can be obtained by tracing Eq. (\ref{eq:rho_id-dt1}) over the bath of the $n$th module and employing the explicit expression for $H_{c,I}(t)$ of Eq. (\ref{eq:hc_it}), which results in 
\ben
&&\frac{\partial}{\partial t}\sigma_{n,I}(t) = -\frac{1}{\hbar^2}\sum_{m\neq n}\int_0^t d\tau  \nonumber \\
&&\hspace{0.41 in}\left (Tr_b\left\{ {\mathcal F}_{nm}(t){\mathcal F}_{mn}(\tau)\rho_{n,I} (\tau) \rho_{bC_n} \right\} \right . \nonumber \\
&&\hspace{0.41 in}+Tr_b\left\{\rho_{bC_n}\rho_{n,I}(\tau) {\mathcal F}_{nm}(\tau) {\mathcal F}_{mn} (t)\right\} \nonumber \\
&&\hspace{0.41 in}-Tr_b\left\{ {\mathcal F}_{nm}(t) \rho_{bC_m}\rho_{m,I}(\tau) {\mathcal F}_{mn}(\tau)  \right\}\nonumber \\
&&\hspace{0.41 in}\left . - Tr_b\left\{{\mathcal F}_{nm} (\tau) \rho_{bC_m}\rho_{m,I} (\tau){\mathcal F}_{mn} (t) \right\} \right )\ ,
\label{eq:kernel-1}
\een
where the fact that $Tr_{bn}Tr_{bCn}=Tr_b$ has been used.  
More informative expressions for  integrands in Eq. (\ref{eq:kernel-1}) can be obtained utilizing the fact that the dynamics of each module under $H_0$ is independent.  For example, in the exciton 
space, the 
matrix elements of the first term can be expressed as
\ben
&&\langle j_n'' |Tr_b \left\{ {\mathcal F}_{nm}(t) {\mathcal F}_{mn}(\tau) \rho_{n,I}(\tau) \rho_{bC_n}\right\} |j_n'''\rangle \nonumber \\
&&=\sum_{j_n,k_m}\sum_{j_n',k_m'} J_{j_nk_m} J_{j_n'k_m'}\nonumber \\
&&\times Tr_{bm}\left\{ \langle k_m| U_m(t,\tau) \rho_{bm} U_m^{g\dagger} (t-\tau)|k_m'\rangle \right\} \nonumber \\
&&\times Tr_{bn} \left \{ \langle j_n'|U_n^g (t-\tau) U_n (\tau) \rho_{n,I}(\tau) |j_n'''\rangle\langle j_n''|U_n^\dagger (t)|j_n\rangle \right\}\ , \nonumber \\ \label{eq:ker1}
\een
where the cyclic invariance of trace operation over the bath of each module has been used.  This expression can be simplified by introducing the following operators of each module  defined in the exciton space: 
\ben 
&&{\mathcal I}_{n} (t,\tau) = Tr_{bn}\left\{ U_n (t,\tau)\rho_{bn} U_n^{g\dagger}(t-\tau)\right\}  \ , \label{eq:an}\\
&&{\mathcal E}_{n,j_n''j_n'''}(t,\tau;\rho_n)=Tr_{bn}\left\{U_n^g(t-\tau)\right . \nonumber \\
&&\hspace{.1 in} \times \left . \left ( \rho_n(\tau) U_n (\tau) |j_n'''\rangle\langle j_n''|U_n^\dagger (\tau) \right ) U_n^\dagger (t,\tau)\right\} \ , \label{eq:en_jj'} 
\een
where $\rho_n(\tau)=U_n(\tau)\rho_{n,I}(\tau)U_n^\dagger (\tau)$.  Then, Eq. (\ref{eq:ker1}) can be expressed as 
\ben
&&\langle j_n'' |Tr_b \left\{ {\mathcal F}_{nm}(t) {\mathcal F}_{mn}(\tau) \rho_{n,I}(\tau) \rho_{bC_n}\right\} |j_n'''\rangle \nonumber \\
&&\hspace{.1in}=\sum_{j_n,k_m}\sum_{j_n',k_m'} J_{j_nk_m} J_{j_n'k_m'}\nonumber \\
&&\hspace{.2in}\times \langle k_m|{\mathcal I}_m (t,\tau) |k_m'\rangle \langle j_n'| {\mathcal E}_{n,j_n''j_n'''}(t,\tau;\rho_n)|j_n\rangle \ .
 \label{eq:ker1c}
\een 
As can be inferred from the definition of Eq. (\ref{eq:en_jj'}), the operator ${\mathcal E}_{n,j_n''j_n'''}(t,\tau;\rho_n)$ is a functional of $\rho_n(\tau)$. 

Similar expressions can be obtained for the other three integrands of Eq. (\ref{eq:kernel-1}).  For these, we introduce two counterparts of Eqs. (\ref{eq:an}) and (\ref{eq:en_jj'}) as follows: 
\ben
&&{\mathcal E}_{n}(t,\tau;\rho_n)=Tr_{bn}\left\{U_n^g(t-\tau)\rho_{n}(\tau)U_n^\dagger (t,\tau)\right\} \ , \label{eq:en} \\ 
&&{\mathcal I}_{n,j_n''j_n'''} (t,\tau) = Tr_{bn}\left\{ U_n (t,\tau) \nonumber \right . \\
&& \hspace{.2in} \left . \times \left (U_n(\tau) |j_n''' \rangle \langle j_n''|U_n^\dagger (\tau) \rho_{bn} \right )U_n^{g\dagger}(t-\tau)\right\} \ . \label{eq:an_jj'}
\een
Then, the time evolution equations for the matrix elements of Eq. (\ref{eq:kernel-1}) can be expressed as
\ben
&&\frac{\partial}{\partial t}\langle j_n''|\sigma_{n,I}(t)|j_n'''\rangle =
-\frac{1}{\hbar^2}\sum_{m\neq n}\sum_{j_n,k_m}\sum_{j'_n,k'_m}J_{j_nk_m}J_{j_n'k_m'}\nonumber \\
&&\times  \int_0^t d\tau \left \{ \langle k_m|{\mathcal I}_m (t,\tau) |k_m'\rangle \langle j_n'| {\mathcal E}_{n,j_n''j_n'''}(t,\tau;\rho_n)|j_n\rangle \right . \nonumber \\ 
&&\hspace{.2in} +\langle k_m'|{\mathcal I}_m^\dagger (t,\tau) |k_m\rangle \langle j_n| {\mathcal E}_{n,j_n'''j_n''}^\dagger(t,\tau;\rho_n)|j_n'\rangle \nonumber \\
&&\hspace{.2in}   - \langle j_n'| {\mathcal I}_{n,j_n'''j_n''}^\dagger (t,\tau)|j_n\rangle\langle k_m|{\mathcal E}_m^\dagger (t,\tau;\rho_m) |k_m'\rangle   \nonumber \\ 
&&\hspace{.2in} \left .  - \langle j_n| {\mathcal I}_{n,j_n''j_n'''} (t,\tau)|j_n'\rangle\langle k_m'|{\mathcal E}_m (t,\tau;\rho_m) |k_m\rangle  \right \} \ .
 \label{eq:sigma_In_dt2}
\een

A time evolution equation for the MED, $p_n(t)=\sum_{j_n}\langle j_n|\sigma_{n,I}(t)|j_n\rangle$, can be obtained by summing the diagonal components of Eq. (\ref{eq:sigma_In_dt2}) and  utilizing the fact that ${\mathcal I}_n(t,\tau)=\sum_{j_n''}{\mathcal I}_{n,j_n''j_n''}(t,\tau)$ and   ${\mathcal E}_n(t,\tau;\rho_n)=\sum_{j_n''}{\mathcal E}_{n,j_n''j_n''}(t,\tau;\rho_n)$, yielding
\ben
&&\frac{\partial}{\partial t} p_n(t) =
-\frac{1}{\hbar^2}\sum_{m\neq n}\sum_{j_n,k_m}\sum_{j'_n,k'_m}J_{j_nk_m}J_{j_n'k_m'}\nonumber \\
&&\times 2 {\rm Re}\int_0^t d\tau \left \{ \langle k_m|{\mathcal I}_m (t,\tau) |k_m'\rangle \langle j_n'| {\mathcal E}_{n}(t,\tau;\rho_n)|j_n\rangle \right . \nonumber \\ 
&&\hspace{.2in} \left .  - \langle j_n| {\mathcal I}_{n} (t,\tau)|j_n'\rangle \langle k_m'|{\mathcal E}_m (t,\tau;\rho_m) |k_m\rangle  \right \}\ .
 \label{eq:sigma_In_dt3}\ 
\een
Higher order versions of 
this equation can be obtained from
Eq. (\ref{eq:rho_id-dt_ex}) 
by following similar procedures including higher than second order terms.
In the limit where each module consists of a single chromophore, the GME-MED reduces to  the GME for localized excitons\cite{kenkre-prb9}.    

Equation (\ref{eq:sigma_In_dt3}) is the main formal result of the present letter, but further simplification is needed for its practical application because of the functional dependence of ${\mathcal E}_n(t,\tau;\rho_n)$ on $\rho_n(t)$.   We now describe a generic approximation that is suitable 
for natural photosynthetic systems and  is also implicit in applications employing MC-FRET rates in the Pauli master equation\cite{ritz-jpcb105,renger-pr102,bennett-jacs}.
To simplify the argument, we shall assume that all $H_n$ are time independent.  
Then $U_n(t,\tau)=U_n(t-\tau)$ and ${\mathcal I}_n(t,\tau)={\mathcal I}_n (t-\tau,0)\equiv{\mathcal I}_n (t-\tau)$.   If the dynamics driving intra-module detailed balance occurs much faster than the inter-module population dynamics, we can make the steady state approximation of $\rho_n(\tau)\approx \rho_n^s p_n(\tau)$, 
where $\rho_n^s=e^{-\beta H_n}/Tr_{n}\{e^{-\beta H_n}\}$.  This does not yet
imply 
{\it complete} time scale separation between intra-module and inter-module dynamics, and takes the full effect of exciton-bath entanglement into consideration through $\rho_n^s$.   With this approximation, ${\mathcal E}_n(t,\tau)\approx {\mathcal E}_n^s (t-\tau)p_n(\tau)$, 
where  ${\mathcal E}_{n}^s(t)=Tr_{bn}\left\{U_n^g(t)\rho_{n}^sU_n^\dagger (t)\right\}$. 
Equation (\ref{eq:sigma_In_dt3}) then reduces to the following closed-form expression:
\ben
&&\frac{\partial}{\partial t} p_n(t) =\sum_{m\neq n}\int_0^t d\tau \left\{ {\mathcal K}_{m\rightarrow n}(t-\tau) p_m(\tau)\right .\nonumber \\
&&\hspace{1.2in}\left . -{\mathcal K}_{n\rightarrow m}(t-\tau) p_n(\tau)\right\} \ ,\label{eq:master_1}
\een 
where
\ben
&&{\mathcal K}_{n\rightarrow m}(t)=\frac{2}{\hbar^2}{\rm Re}\sum_{j_n,k_m}\sum_{j_n',k_m'}J_{j_nk_m}J_{j_n'k_m'} \nonumber \\
&&\hspace{.7in}\times \langle k_m|{\mathcal I}_m (t) |k_m'\rangle \langle j_n'| {\mathcal E}_{n}^s(t)|j_n\rangle \ . \label{eq:k_nmt_gme}
\een
The GME-MED of Eq. (\ref{eq:master_1}) now can be solved employing the pre-determined kernels of Eq. (\ref{eq:k_nmt_gme}),  which can be evaluated using appropriate lineshape theories. Alternatively, a time-local version of Eq. (\ref{eq:master_1}) can be obtained by replacing $p_m(\tau)$ and $p_n(\tau)$ in the integrand with $p_m(t)$ and $p_n(t)$, respectively.   When all the intra-module exciton dynamics are much faster than the inter-module dynamics, the assumption of complete time scale separation reduces Eq. (\ref{eq:k_nmt_gme}) to the Pauli master equation with time independent transition rate, $\tilde {\mathcal K}_{n \rightarrow m}=\int_0^\infty dt\ {\mathcal K}_{n\rightarrow m}(t)$.  
This further becomes identical to the MC-FRET rate\cite{jang-prl92} when expressed in terms of overlap of lineshape functions $I_m^{k_mk_m'}(\omega)=\int_{-\infty}^\infty d t\ e^{i\omega t} \langle k_m|{\mathcal I}_m(t)|k_m'\rangle$ and $E_n^{j_n'j_n}(\omega)=2\ {\rm Re} \int_0^\infty dt \ e^{-i\omega t} \langle j_n'|{\mathcal E}_n^s(t)|j_n\rangle$.

As a demonstration,  we consider a system consisting of bacteriochlorophylls (BChls) 1-4 
in the Fenna-Matthews-Olson (FMO) complex and its protein bath,  using parameters adopted from previous works\cite{ishizaki-pnas106,hoyer-pre86}.   We model this as a two-module system (Fig. 2 (a)). The exciton Hamiltonian of each module is given by $H_n^e=E_{1_n}|1_n\rangle\langle 1_n|+E_{2_n}|2_n\rangle\langle 2_n|+\Delta_n(|1_n\rangle\langle2_n|+|2_n\rangle \langle 1_n|)$, for $n=1,2$. The bath is modeled as a site-local Ohmic-Drude bath\cite{ishizaki-pnas106,hoyer-pre86} with reorganization energy of $\lambda=35\ {\rm cm^{-1}}$ and Drude cutoff at $\hbar \omega_c=106 \ {\rm cm^{-1}}$.   
Accurate calculations are first made with the hierarchical equation of motion (HEOM) approach\cite{ishizaki-jcp130-2}, which is known to be virtually exact  for  this spectral density.  The resulting modular total excitonic densities calculated for two different initial conditions, one starting from $|1_1\rangle$ and the other starting from $|2_1\rangle$ are shown in Fig. 2 as blue and red dashed lines, respectively, at 
$T=150$ and $300 \ {\rm K}$.  Although the population at each BChl is sensitive to the initial condition and exhibits strongly coherent behavior (see insets), the modular excitonic density shows monotonic behavior and is much more insensitive to the initial condition.  

\begin{figure}
(a)\makebox[1.1in]{}(b)\makebox[2.1in]{ }\\
\includegraphics[width=1.1 in]{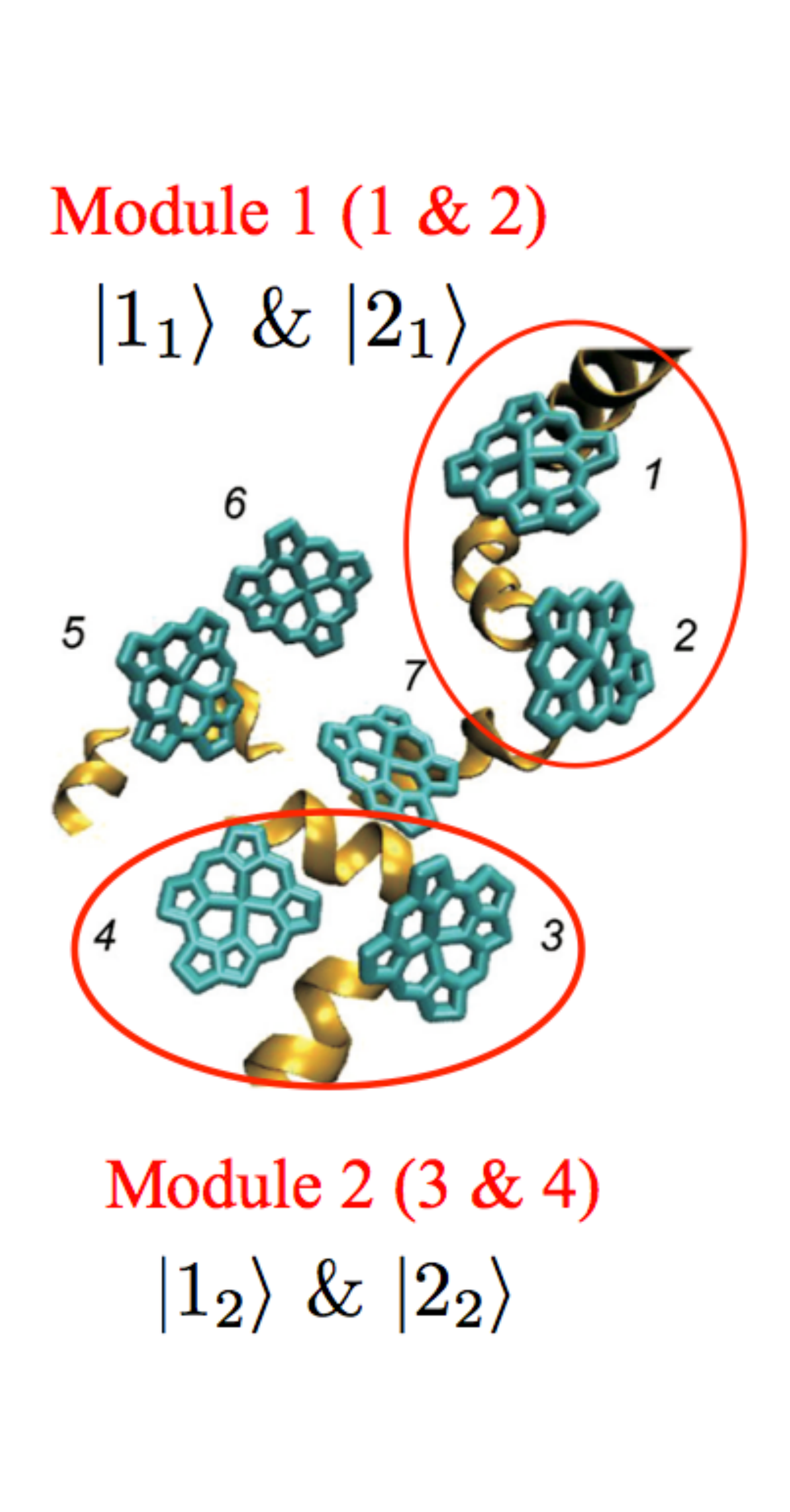}\hspace{.1in}\includegraphics[width=2.1in]{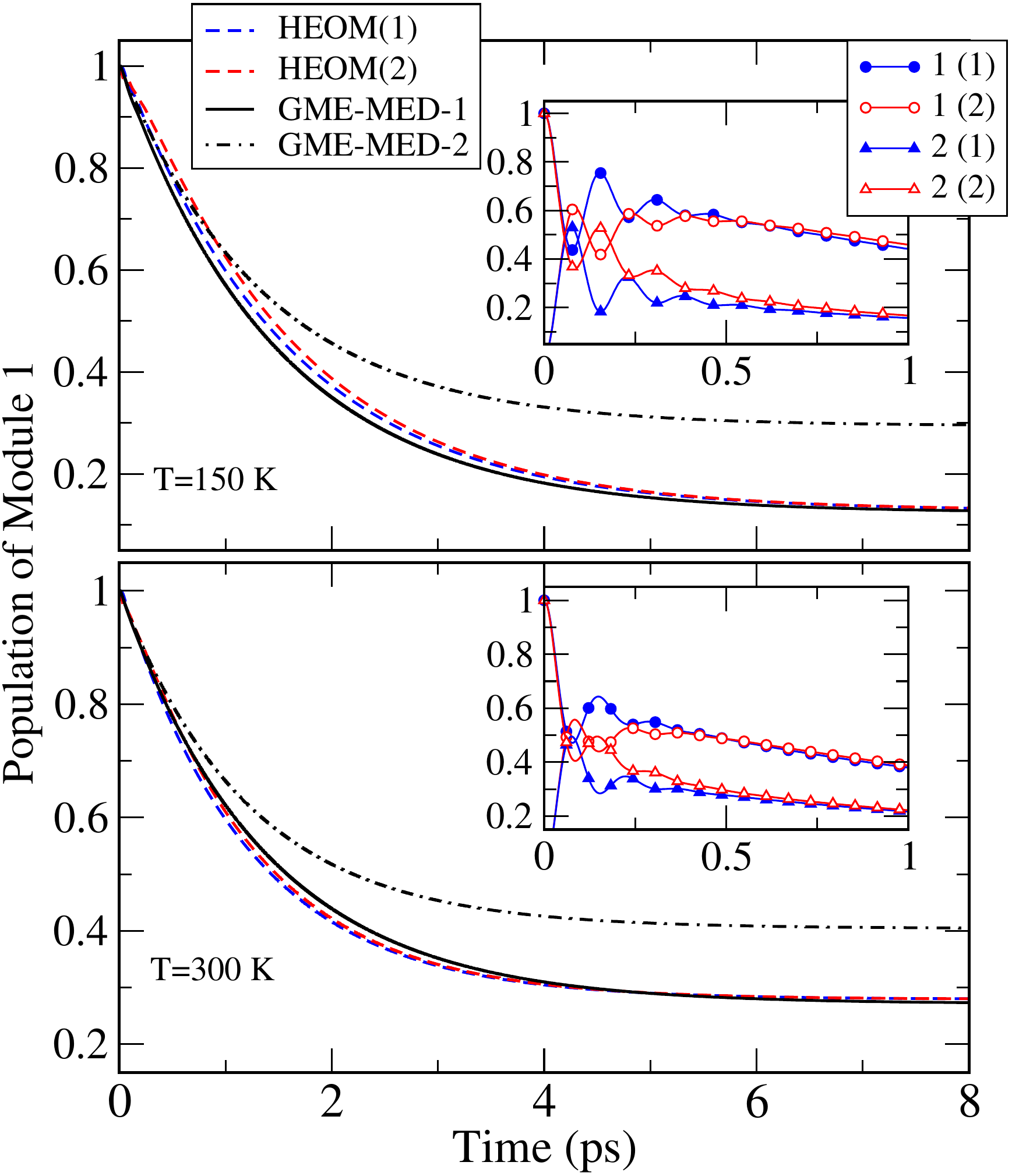}
\caption{(a) Decomposition of first four BChls of FMO complex into two modules.  
The parameters defining $H_{n}^e$ and $H_c$ (all in ${\rm cm^{-1}}$) are: $E_{1_1}=12,400$; $E_{2_1}=12,520$; $E_{1_2}=12,200$; $E_{2_2}=12,310$; $\Delta_1=-87$; $\Delta_2=-53$; $J_{1_11_2}=5$; $J_{1_12_2}=-5$; $J_{2_11_2}=30$; $J_{2_12_2}=8$. (b) Time dependent populations of module 1 calculated with HEOM and with two different approximations for GME-MED.  Insets show HEOM populations at each BChl.   In all figures, numbers within parentheses represent the site of initial excitation.  }
\end{figure}

Employing the time-local version of Eq. (\ref{eq:master_1}), which still accounts for non-Markovian effects,
the time dependent MED was then calculated with two approximations for  Eq. (\ref{eq:k_nmt_gme}) as described below.  Denote the eigenstate $H_n^e$ of module $n$ with energy $\epsilon_{p_n}$ as $|\varphi_{p_n}\rangle$, and define the unitary transformation  matrix element as $U_{j_np_n}=\langle j_n|\varphi_{p_n}\rangle$.   Then, neglecting the off-diagonal elements of exciton-bath couplings in the exciton basis, and employing the following lineshape function for the Ohmic-Drude spectral density: $g_{\lambda}(t)=\frac{2\lambda}{\beta\hbar^2 \omega_c} t +\frac{\lambda}{\hbar\omega_c} \cot(\frac{\beta\hbar\omega_c}{2})\left ( e^{-\omega_c t}-1\right) +\frac{4\lambda \omega_c}{\beta\hbar^2} \sum_{l=1}^\infty \frac{e^{-\omega_l t}-1}{\omega_l (\omega_l^2-\omega_c^2)}+i\frac{\lambda}{\hbar\omega_c} (1-e^{-\omega_c t})$,
with $\omega_l=2\pi l/(\beta \hbar)$, we can express Eq. (\ref{eq:k_nmt_gme}) as 
\ben 
&&{\mathcal K}_{n\rightarrow m} (t)\approx \frac{2}{\hbar^2} {\rm Re} \sum_{p_n,p_m'} \frac{e^{-\beta \tilde \epsilon_{p_n}} }{(\sum_{p_n''} e^{-\beta \tilde \epsilon_{p_n''}})} |\tilde J_{p_np_m'}|^2 \nonumber \\
&&\hspace{0.4 in}\times  e^{-g_{\lambda_{p_n}}(t)-g_{\lambda_{p_m'}}(t)+i(\tilde \epsilon_{p_n}-\tilde \epsilon_{p_m'})t/\hbar} \ , \label{eq:knm-app1}
\een
where $\tilde J_{p_np_m'}=\sum_{j_n,k_m} U_{k_mp_m'} J_{j_nk_m} U^*_{j_np_n}$\cite{notedifferenceICC}, 
$\lambda_{p_n}=(\sum_{j_n} |U_{j_n,p_n}|^4 )\lambda$, and $\tilde \epsilon_{p_n}=\epsilon_{p_n}-\lambda_{p_n}$.   Equation (\ref{eq:knm-app1}) incorporates all orders of exciton-phonon coupling.  The corresponding results are denoted as GME-MED-1 in Fig. 2(b) and are seen to show excellent agreement with the corresponding HEOM populations  both in the initial times and the steady state limits. In the second approximation, denoted as GME-MED-2 in Fig. 2(b), we employ approximate values of ${\mathcal I}_m (t)$ and ${\mathcal E}_{n}(t)$ calculated by the 2nd order time-local quantum master equation approach\cite{jang-jcp118-1} and neglecting exciton-bath entanglement in the initial state of the emission lineshape function.  Fig. 2(b) shows that this results in a less accurate representation.

Considering the simplicity of (\ref{eq:knm-app1}), the good agreement between the results of GME-MED-1 and HEOM at both low and room temperature  is surprising. 
It suggests that the net contribution of non-equilibrium effects, 
inter-module non-adiabatic couplings and quantum coherence, which are not fully accounted for in this approximation,  have relatively  minor contributions to the dynamics of MED in this system.  Indeed, comparison with the results in the Markovian limit (not shown) confirmed that non-Markovian effects are not significant in this system. On the other hand, 
relative poor performance of GME-MED-2, which neglects the exciton-phonon couplings beyond the second  order and of the initial exciton states,   shows 
that inclusion of all the higher order exciton-bath coupling is crucial for obtaining correct steady state limits. These results suggest that master equation approaches \cite{ritz-jpcb105,novoderezhkin-pccp13,renger-pr102,renger-jpp168,bennett-jacs} may attain reliable accuracy for large scale systems provided that proper division of the system into appropriate modules and use of accurate lineshape functions is made.    Analysis of these issues for real large scale light harvesting complexes such as PSII \cite{bennett-jacs} and the light harvesting apparatus of green sulfur bacteria\cite{fujita-jpcl3,huh-arxiv1307.0886} can be made by comparison of GME-MED with high level calculations for small subsets as demonstrated here for FMO complex 
or even for medium size systems\cite{strumpfer-jcp137,hein-njp14,kreisbeck-jpcl3}, 
and by comparing results based on different levels of lineshape theory \cite{jang-jcp118-1,mukamel,May}. 

In summary, we have presented a general derivation of a 
generalized master equation for coherent excitonic energy transfer between modules of chromophores. As a proof of principle demonstration we showed
that this approach allows the coherent population dynamics in sub-complexes of FMO
to be accurately described by transitions between modular exciton densities, opening a novel route to calculation of long range transfer of excitonic energy between modules within which electronic coherence contributes.

This work was supported by DARPA under Award No.\ N66001-09-1-2026. SJ also acknowledges support by the National Science Foundation CAREER award (Grant No. CHE-0846899), the Office of Basic Energy Sciences, Department of Energy (Grant No. DE-SC0001393), and the Camille Dreyfus Teacher Scholar Award.  SH is a DOE Office of Science Graduate Fellow.  GRF also acknowledges support from the Director, Office of Science, Office of Basic Energy Sciences, of the US Department of Energy under contract DE-AC02-05CH11231.
We thank H. Choe for rendering the image of Fig. 1.


\begin{thebibliography}{10}

\bibitem{hu-qrb35}
{X. Hu, T. Ritz, A. Damjanovic, F. Autenrieth, and K. Schulten}, Quar. Rev.
  Biophys. {\bf 35},  1  (2002).

\bibitem{blankenship-ps}
R.~E. Blankenship, {\em Molecular Mechanism of Photosynthesis} (Blackwell
  Science, Oxford, UK, 2002).

\bibitem{jang-jpcb111}
S. Jang, M.~D. Newton, and R.~J. Silbey, J. Phys. Chem. B {\bf 111},  6807
  (2007).

\bibitem{rebentrost-njp11}
{P. Rebentrost, M. Mohseni, I. Kassal, S. Lloyd, and A. Aspuru-Guzik}, New J.
  Phys. {\bf 11},  033003  (2009).

\bibitem{caruso-pra81}
{F. Caruso, A. W. Chin, A. Datta, S. F. Huelga, and M. B. Plenio}, Phys. Rev. A
  {\bf 81},  062346  (2010).

\bibitem{wu-njp12}
{J. L. Wu, F. Liu, Y. Shen, J. S. Cao, R. J. Silbey}, New J. Phys. {\bf 12},
  105012  (2010).

\bibitem{ishizaki-pnas106}
A. Ishizaki and G.~R. Fleming, Proc. Natl. Acad. Sci., USA {\bf 106},  17255
  (2009).

\bibitem{huo-jpcl2}
P. Huo and D.~F. Coker, J. Phys. Chem. Lett. {\bf 2},  825  (2011).

\bibitem{strumpfer-jcp137}
{J. Str\"{u}mpfer and K. Schulten}, J. Chem. Phys. {\bf 137},  065101  (2012).

\bibitem{hein-njp14}
B. Hein, C. Kreisbeck, T. Kramer, and M. Rodriguez, New. J. Phys. {\bf 14},
  023018  (2012).

\bibitem{kreisbeck-jpcl3}
C. Kreisbeck and T. Kramer, J. Phys. Chem. Lett. {\bf 3},  2828  (2012).

\bibitem{ritz-jpcb105}
T. Ritz, S. Park, and K. Schulten, J. Phys. Chem. B {\bf 105},  8259  (2001).

\bibitem{sener-jcp120}
{M. K. Sener, S. Park, D. Lu, A. Damjanovi\'{c}, T. Ritz, P. Fromme, and K.
  Schulten}, J. Chem. Phys. {\bf 120},  11183  (2004).

\bibitem{yang-bj85}
{M. Yang, A. Damjanovi\'{c}, H. M. Vaswani, and G. R. Fleming}, Biophys. J.
  {\bf 85},  140  (2003).

\bibitem{novoderezhkin-pccp13}
{V. I. Novoderezhkin, A. Marin, and R. van Grondelle}, Phys. Chem. Chem. Phys.
  {\bf 13},  17093  (2011).

\bibitem{renger-pr102}
T. Renger, Photosyn. Res. {\bf 102},  471  (2009).

\bibitem{renger-jpp168}
{T. Renger, M. E. Madjet, A. Knorr, and F. M\"{u}h}, J. Plant Physiol. {\bf
  168},  1497  (2011).

\bibitem{bennett-jacs}
D.~I.~G. Bennett, K. Amarnath, and G.~R. Fleming, J. Am. Chem. Soc. {\bf 135},
  9164  (2013).

\bibitem{hoyer-pre86}
S. Hoyer, A. Ishizaki, and K.~B. Whaley, Phys. Rev. E {\bf 86},  041911
  (2012).

\bibitem{scholes-jpcb105}
G.~D. Scholes, X.~J. Jordanides, and G.~R. Fleming, J. Phys. Chem. B {\bf 105},
   1640  (2001).

\bibitem{jang-prl92}
S. Jang, M.~D. Newton, and R.~J. Silbey, Phys. Rev. Lett. {\bf 92},  218301
  (2004).

\bibitem{vankampen-jsp87}
{N. G. van Kampen and I. Oppenheim}, J. Stat. Phys. {\bf 87},  1325  (1997).

\bibitem{jang-jcp116}
S. Jang, J. Cao, and R.~J. Silbey, J. Chem. Phys. {\bf 116},  2705  (2002).

\bibitem{renger-pr343}
T. Renger, V. May, and O. K\"{u}hn, Phys. Rep. {\bf 343},  137  (2001).

\bibitem{miller-jpca105}
W.~H. Miller, J. Phys. Chem. A {\bf 105},  2942  (2001).

\bibitem{makri-arpc50}
N. Makri, Annu. Rev. Phys. Chem. {\bf 50},  167  (1999).

\bibitem{kenkre-prb9}
V.~M. Kenkre and R.~S. Knox, Phys. Rev. B {\bf 9},  5279  (1974).

\bibitem{ishizaki-jcp130-2}
A. Ishizaki and G.~R. Fleming, J. Chem. Phys. {\bf 130},  234111  (2009).

\bibitem{notedifferenceICC}
Note that this coupling between excitons in different modules differs from the
  inter-complex coupling (ICC) of Ref. \cite{hoyer-pre86}.

\bibitem{jang-jcp118-1}
S. Jang and R.~J. Silbey, J. Chem. Phys. {\bf 118},  9312  (2003).

\bibitem{fujita-jpcl3}
{T. Fujita, J. C. Brookes, S. K. Saikin, and A. Aspuru-Guzik}, J. Phys. Chem.
  Lett. {\bf 3},  2357  (2012).

\bibitem{huh-arxiv1307.0886}
J. Huh {\it et~al.}, arXiv:1307.0886  (2013).

\bibitem{mukamel}
S. Mukamel, {\em Principles of Nonlinear Spectroscopy} (Oxford University
  Press, New York, 1995).

\bibitem{May}
{Volkard May and Oliver K\"{u}hn}, {\em Charge and Energy Transfer Dynamics in
  Molecular Systems} (Wiley-VCH, Weinheim, Germany, 2011).

\end{thebibliography}
\end{document}